\documentclass[superscriptaddress,twocolumn,amsmath,amssymb]{revtex4-1}

\usepackage{amsmath,amssymb}
\usepackage{graphicx}
\usepackage{verbatim}
\usepackage{color}
\usepackage[english]{babel}
\usepackage{bm}
\usepackage{natbib}
\usepackage[symbol]{footmisc}
\usepackage{textcomp}
\newcommand{\red}{\textcolor{red}}

\usepackage{setspace}

\begin{document}

\title{Single-particle Mie-resonant all-dielectric nanolasers}

\author{Ekaterina Yu.~Tiguntseva}
\affiliation{ITMO University, St.~Petersburg 197101, Russia}

\author{Kirill L.~Koshelev}
\affiliation{ITMO University, St.~Petersburg 197101, Russia}
\affiliation{Nonlinear Physics Center, Australian National University, Canberra ACT 2601, Australia}

\author{Aleksandra D.~Furasova}
\affiliation{ITMO University, St.~Petersburg 197101, Russia}

\author{Vladimir Yu.~Mikhailovskii}
\affiliation{St.~Petersburg State University, St. Petersburg 199034, Russia}

\author{Elena V.~Ushakova}
\affiliation{ITMO University, St.~Petersburg 197101, Russia}

\author{Denis G.~Baranov}
\affiliation{Department of Physics, Chalmers University of Technology, 412 96 Gothenburg, Sweden} 

\author{Timur O.~Shegai}
\affiliation{Department of Physics, Chalmers University of Technology, 412 96 Gothenburg, Sweden} 

\author{Anvar A.~Zakhidov}
\affiliation{ITMO University, St.~Petersburg 197101, Russia}
\affiliation{University of Texas at Dallas, Richardson, TX 75080, USA}

\author{Yuri S.~Kivshar}
\affiliation{ITMO University, St.~Petersburg 197101, Russia}
\affiliation{Nonlinear Physics Center, Australian National University, Canberra ACT 2601, Australia}
\email{ysk@internode.on.net}

\author{Sergey V. Makarov}
\affiliation{ITMO University, St.~Petersburg 197101, Russia}

\begin{abstract}
All-dielectric subwavelength structures utilizing Mie resonances provide a novel paradigm in nanophotonics for controlling and manipulating light~\cite{kuznetsovreview}. So far, only spontaneous emission enhancement was demonstrated with single dielectric nanoantennas ~\cite{rutckaia2017quantum, tiguntseva2018light}, whereas stimulated emission was achieved only in large lattices supporting collective modes~\cite{ha2018directional}. Here, we demonstrate the first single-particle all-dielectric monolithic nanolaser driven by Mie resonances in visible and near-IR frequency range. We employ halide perovskite CsPbBr$_3$ as both gain and resonator material that provides high optical gain (up to $\sim$10$^4$ cm$^{-1}$)  and allows simple chemical synthesis of nanocubes with nearly epitaxial quality~\cite{kelso2019spin}. Our smallest non-plasmonic Mie-resonant single-mode nanolaser with the size of 420-nm operates at room temperatures and wavelength 535~nm with linewidth $\sim 3.5$ meV.
%It is characterized by the lasing threshold up to 1.5~GW/cm$^2$ and  $\sim 3.5$ meV. 
These novel lasing nanoantennas can pave the way to multifunctional photonic designs for active control of light at the nanoscale.
\end{abstract}
 
\maketitle

All-dielectric nanoantennas employing Mie resonances with weak losses and low-damage threshold in visible and near-infrared spectral range provide new functionalities to tailor the properties of light at the nanoscale not available with the use of plasmonics~\cite{kuznetsovreview}. Moreover, this novel platform allows strong active response from the nanoparticle material, resulting in the enhanced generation of nonlinear harmonics~\cite{shcherbakov2014enhanced, kruk2019nonlinear}, Raman scattering~\cite{dmitriev2016resonant}, and photoluminescence~\cite{rutckaia2017quantum, tiguntseva2018light}. Recent results for stimulated emission from individual dielectric nanoparticles are limited by numerical simulations~\cite{fratalocchi2008three, gongora2017anapole}, and only large arrays of resonant nanoparticles have been employed for the experimental studies of stimulated emission~\cite{ha2018directional}. Relatively low $Q$-factors of resonant modes ($Q_m$) supported by single dielectric nanoparticles still prevent the development of all-dielectric nanolasers with subwavelength dimensions, and they require  materials with a large gain.

It was realized very recently that one of the best materials for lasing nanoantenna is metal-halide perovskites. Their high-enough refractive index (higher than 2) makes it possible to develop a compact design~\cite{sutherland2016perovskite, makarov2019halide, gao2018lead}. Also, most of halide perovskites support excitons at room temperatures~\cite{su2018room} and, thus, high density of states near the bottom of conduction band can yield high luminescence quantum efficiency and optical gain~\cite{sutherland2016perovskite}. Finally, relatively simple chemical methods for the fabrication of optically resonant structures with regular shapes (nanowires~\cite{zhu2015lead, xing2015vapor, shang2018surface}, microplates~\cite{zhang2014room, zhang2016high, su2017room}, and microspheres~\cite{tang2017single}) allow generating stimulated emission in the range of $420 \div 824$~nm~\cite{zhu2015lead, fu2016nanowire, eaton2016lasing, park2016light}. However, only cubic particles~\cite{zhou2018single} can result in the fabrication of truly subwavelength single-particle nanoantennas with low defects that can support lasing at room temperatures.

\begin{figure*}
  \includegraphics[width=0.9\linewidth]{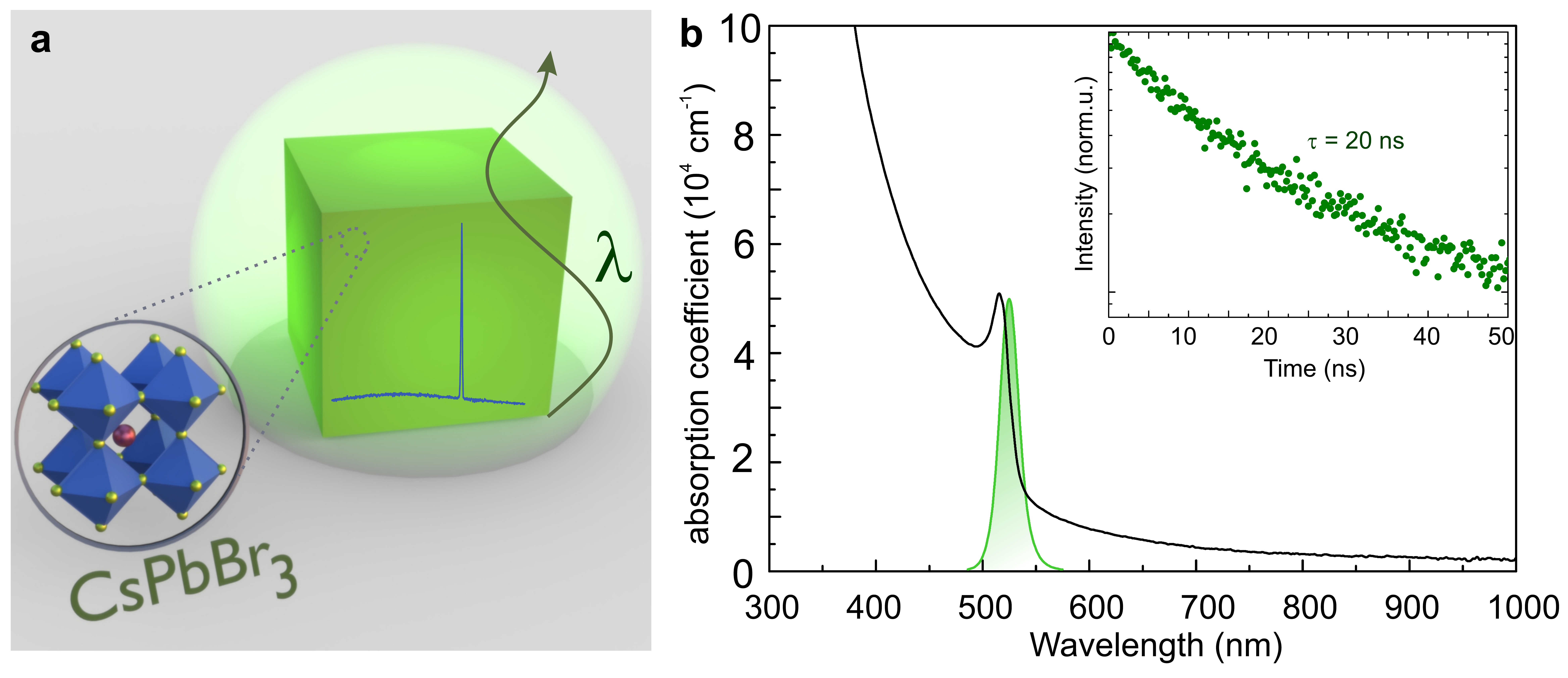}
  \caption{{\bf Single-particle Mie nanolaser.} (a) General concept. (b) Material properties of CsPbBr$_3$ perovskite: absorption spectrum of a thin film (black curve) and photoluminescence spectrum (green curve). Inset shows time-resolved photoluminescence for a typical perovskite nanocube grown on a sapphire substrate.}
  \label{fig1}
\end{figure*}

%\textbf{Fabrication and characterization}
 Figure~\ref{fig1} illustrates a nanolaser made of CsPbBr$_3$ being a material with high direct bandgap absorption ($\alpha>$~10$^4$~cm$^{-1}$) and high density of states at the edge of the conduction band, allowing an overlap with a photoluminescence (PL) peak centered at $\lambda$=532~nm with the full-width at half maximum (FWHM) of 14~nm. Despite high optical losses at the PL wavelengths, they are converted to an useful gain once the photoexcitation level becomes high enough to fill the levels at the excitonic state. 
 
\begin{figure*}
  \includegraphics[width=1\linewidth]{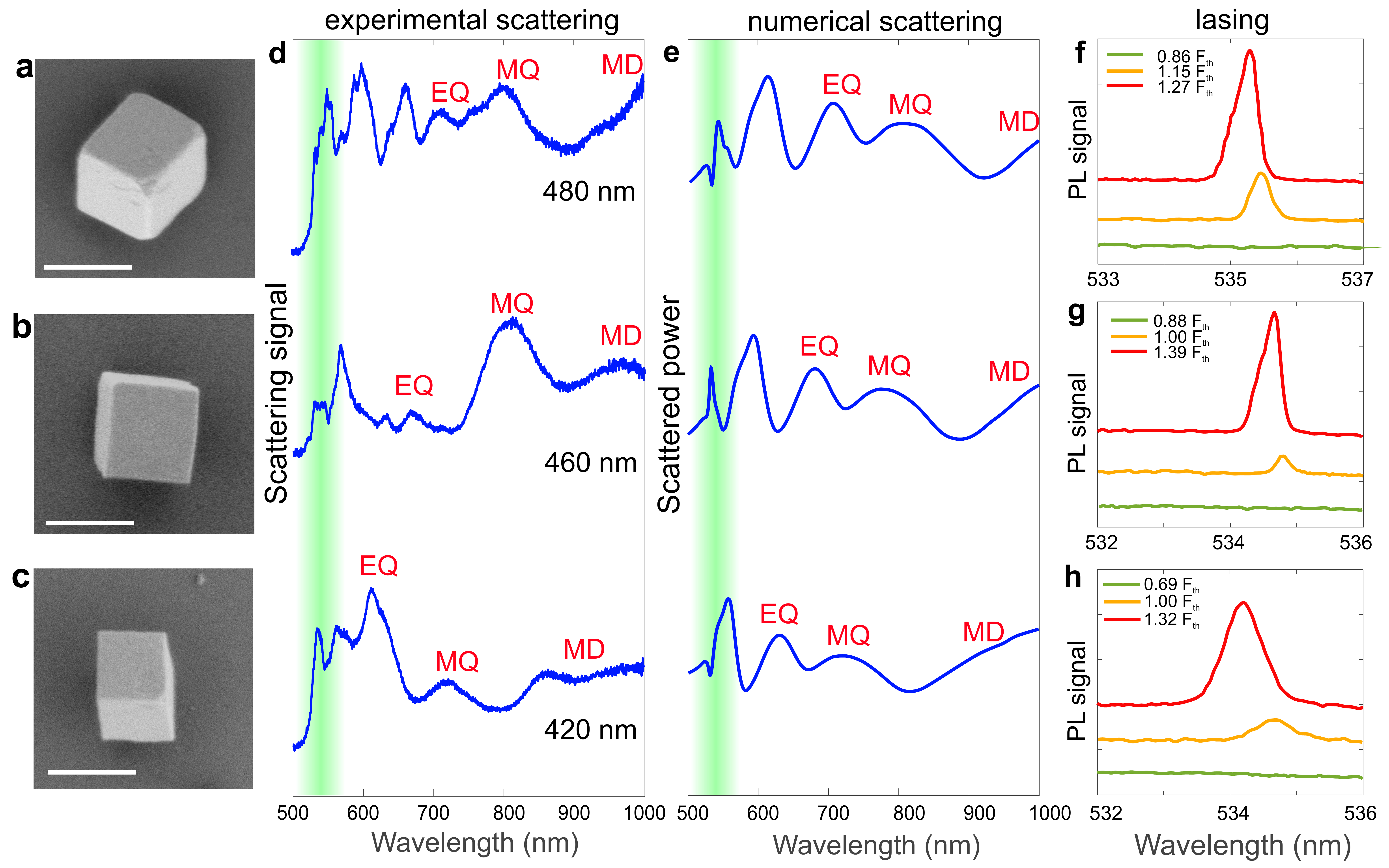}
  \caption{\textbf{Optical properties of resonant CsPbBr$_3$ nanocubes.} (a-c) SEM images of nanocubes placed on a sapphire substrate. Scale bar is 500~nm. (d) Experimental and (e) theoretical dark-field spectra of the CsPbBr$_3$ nanocubes for the $s$-polarized incident white light. (f-h) Pump intensity-dependent emission spectra at different values of fluence below and above the lasing threshold, for the selected nanocubes.}
  \label{fig2}
\end{figure*}

CsPbBr$_3$ nanocubes are synthesized chemically on a sapphire substrate by a two-step deposition method (see \textit{Methods}). The time-resolved PL demonstrates an exponential decay with an average lifetime of 20~ns homogeneous over whole nanoparticle, being also confirmed by the time-resolved PL map [see Fig.~\ref{fig1}(b) and Fig.~\red{S3}]. Indeed, CsPbBr$_3$ is a defect-tolerant material where defect sates are placed very close to the conduction or valence bands keeping the bandgap clean~\cite{brandt2017searching}. Nevertheless, for the nanocubes grown on fused silica, lasing occurs for those larger than the emission wavelength only, caused by worse material properties (for details, see Supplementary Section~\red{S3}). The nanoparticles are characterized by means of scanning electron microscopy (SEM). As can be seen in Figs.~\ref{fig2}(a-c), the predominant aspect ratio of obtained nanostructures is close to 1:1:1, whereas the quality of facets is high enough to support high-$Q$ resonant optical modes.

Optical resonances of individual nanocubes with different sizes are studied with dark-field (DF) spectroscopy (for details, see \textit{Methods}). The DF spectra are measured for a set of subwavelength nanocubes, as shown in Fig.~\ref{fig2}(a-c), and they represent the behaviour typical for all-dielectric nanoantennas where low-order Mie-type resonances strongly affect the scattering properties, which are employed for the manipulation of light at the subwavelength scale with conventional semiconductors~\cite{kuznetsovreview}. In order to reveal the origin of the observed resonances, we employ numerical simulations (using COMSOL Multiphysics) of the plane wave scattered by a perovskite nanocube placed on a substrate, and illuminated from a certain angle that corresponds to our experimental conditions (for details, see \textit{Methods}). The simulated scattering spectra are combined with the multipole decomposition (for details, see Supplementary Section~\red{S4}), and they reveal the importance of the magnetic dipole (MD), electric (EQ) and magnetic (MQ) quadrupolar modes in both visible and near-IR frequency ranges shifting from $\lambda=700 \div 800$~nm to $\lambda=900$~nm with an increase of the nanocube size [see Figs.~\ref{fig2}(d,e)]. 

The PL spectra of the nanocubes upon excitation with 150~fs laser pulses (for other details, see \textit{Methods}) are presented in Figs.~\ref{fig2}(f-h). For all sizes of the nanocubes, the spectra exhibit threshold-like appearance of a narrow peak on the red side of the emission (exciton) line, in the range of $\lambda=532 \div 538$~nm. The linewidths (FWHM) of these peaks vary from one particle to another, and it corresponds to $Q=670 \div 1600$. The nanocubes can operate in the lasing regime during 10$^6$ excitation cycles in air, whereas after 4$\times$10$^6$ cycles the lasing power looses a half of its initial value, and saturates around that value (see Fig.~\red{S8}). 

\begin{figure*}
  \includegraphics[width=1.0\linewidth]{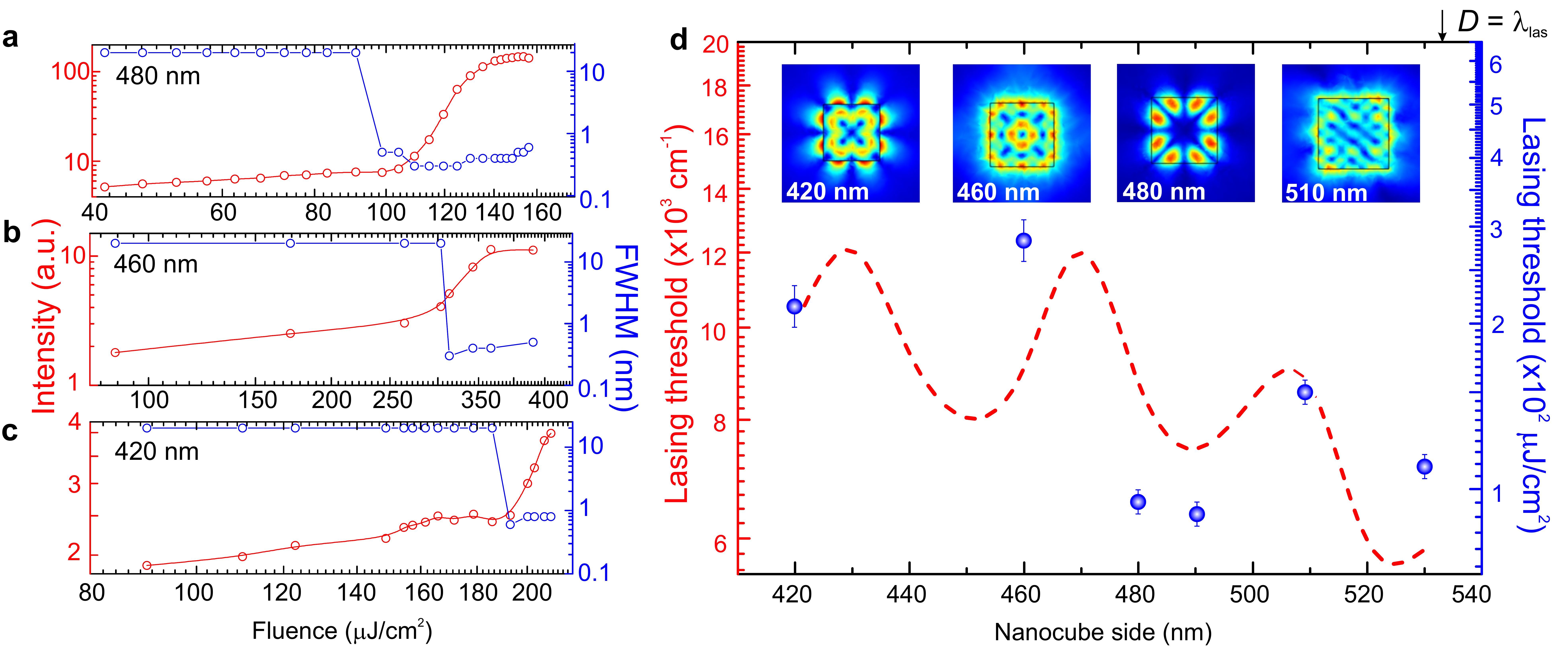}
  \caption{\textbf{Lasing characteristics.} Output emission intensity (red) and FWHM (blue) with increasing pump fluence for 480~nm (a), 460~nm (b), and 420~nm (c) perovskite nanocubes. (d) Experimental dependence of the threshold pump fluence (blue points) and theoretical dependence of threshold gain (red curve) vs. nanocube size. Insets: near-field profiles of the corresponding eigenmodes for the nanocubes of various sizes.}
  \label{fig3}
\end{figure*}

%\textbf{Lasing threshold.} 
To corroborate that the observed emission can be attributed to lasing, we study the input-output power dependence of the system. All the nanocubes demonstrate a recognizable $S$-shape dependence of the emission intensity on the pump power, and a spectral narrowing with increasing pump intensity, as shown in Figs.~\ref{fig3}(a-c). These observations provide strong indications of the lasing regime in our system. Moreover, optical images of the nanocubes taken below and above the lasing threshold further indicate coherence of the emitted light owing to the appearance of interference fringes (see Fig.\red{S9}). 

Additionally, the obtained $P_{\rm in}-P_{\rm out}$ curves for the nanocubes provide useful information about the coupling of spontaneous emission to a lasing mode (beta factor, $\beta$), being an important parameter in the set of laser rate equations (see Supplementary Section~\red{S7}). According to the experimentally measured ratios of stimulated and spontaneous emission, we can extract the following values: $\beta$(480~nm)~$\approx$~0.07, $\beta$(460~nm)~$\approx$~0.33, and $\beta$(420~nm)~$\approx$~0.62. Such a trend is typical for plasmonic nanolasers where $\beta	\rightarrow$~1 for deeply subwavelength nanolasers~\cite{gu2013purcell}, being a general strategy to achieve no-threshold lasing~\cite{ma2019applications}.

Further insight into the observed lasing behavior is provided by Fig.~\ref{fig3}(d), which presents more detailed dependence of the measured threshold pump fluences on the side length of the nanocube. This dependence is supported by theoretical calculations of the threshold gain (expressed in cm$^{-1}$) for perovskite nanocubes of different sizes placed on a sapphire substrate, calculated with the finite-difference time-domain simulations. The material of the nanocubes is modeled by a constant permittivity with a single negative Lorentzian providing gain (for details, see \textit{Methods}). The lasing threshold is then found as singularity of the scattering spectrum, in accordance with the linear theory of lasing~\cite{PhysRevA.82.063824}.

We observe that the experimental data and theoretical results are in a good agreement (we notice that they display thresholds of different quantities -- threshold pump fluence and threshold material gain, respectively). Overall, the lasing threshold decreases with the nanocube size. This decrease is caused mostly by fast increase of $Q$ factors of the nanocube optical modes with its size (see discussion in Supplementary Section~\red{S7}). Smaller oscillations of the threshold gain superimposed on the overall attenuation originate from a finite linewidth of the emitting transition of the gain medium. Depending on the nanocube size, the lasing mode can match exactly the frequency of the gain medium resulting in a reduced threshold. However, when the lasing mode is detuned from the gain medium, the threshold gain increases, also causing the frequency pulling effect~\cite{PhysRevA.82.063824}.

%\textbf{Eigenmode analysis.} 
To analyze the mode structure of the record-small nanolaser, we calculate the eigenmode spectrum of a $420$-nm side nanocube (see \textit{Methods}). Figure~\ref{fig4}(a) shows the $Q$ factors and wavelengths of the eigenmodes in the range of $525 \div 545$~nm. For a nanocube in free space, only one triply-degenerate mode with $Q\approx 70$ can be identified in this spectral range. For a nanocube placed on a sapphire substrate ($n=1.76$), the mode degeneracy is partially lifted, and the triply-degenerate mode splits into one non-degenerate high-$Q$ mode ($Q_{m,v}=30$) and one doubly-degenerate low-$Q$ mode ($Q_{m,h}=15$, where the indices $v$ and $h$ stand for the vertical and horizontal directions). The difference in $Q$ factors of two modes explains the observed single-mode lasing. To characterize the near-field structure of the modes, we perform the multipole decomposition (see \textit{Methods}), and the results are shown in Fig.~\ref{fig4}(b). The electromagnetic field of all three modes is dominated by the electric multipoles of the fourth-order (hexadecapole), and this is confirmed by the near-field patterns presented in Fig.~\ref{fig4}(c).

In conclusion, we have demonstrated active dielectric Mie-resonant nanoantenna for efficient light manipulation in visible/near-IR ranges and  single-mode lasing at room temperatures. The design employs resonant states of Mie-like modes of nanocubes corresponding to the subwavelength regime (down to 0.5$\lambda^3$). As a material providing a high gain ($\sim$10$^4$cm$^{-1}$) we have employed halide perovskite (CsPbBr$_3$) allowing a simple chemical synthesis of nanocubes and supporting moderate-threshold pulsed lasing (1.5~GW/cm$^2$) with high $Q$-factor ($Q>670$) at room temperatures. Table~\red{S1} in Supporting information compares the proposed design with previously reported nanolasers made of conventional semiconductors (such as InGaAs, InGaP, and ZnO), and demonstrates that the perovskite nanolasers hold the record small volume among non-plasmonic nanolasers operating at room temperatures. The developed platform may be applied not only in established applications of nanolasers~\cite{ma2019applications}, but also in advanced structural coloring where Mie resonances are used for both active and passive colors generation~\cite{gao2018lead}.

\section*{Methods}

\textbf{Fabrication of perovskite nanolasers.} The perovskite CsPbBr$_3$ nanocubes are synthesized by a two-step deposition technique. First, sapphire substrates are washed with de-ionized water, and then they are placed into an ozonation station for 20 min. After the dissolving in dimethylformamide (DMF) at 75~$^o$C 0.17~mM PbBr$_2$, the prepared solution is spin-coated on the sapphire substrate at 2500~rpm for 40~s and dried on a hot plate at 75~$^o$C for 30~min. Then, the substrates are dipped into a 7.5~mg/mL CsBr solution in methanol at 50~$^o$C for 15~min. The obtained thin films are rinsed with anhydrous 2-propanol and calcined at 150~$^o$C for 30~min to obtain CsPbBr$_3$ nanocubes. Importantly, this protocols allowed us to avoid deformed shapes of nanocubes, when they are too close to each other~\cite{liu2018robust} and it is hard to study optical properties of a single nanoparticle.

\textbf{Shape and size characterization.}  The perovskite cubes are visualized and inspected by low-voltage scanning electron microscopy with Zeiss Auriga FIB-SEM station at accelerating voltage 700~V and probe current 200~pA. The geometrical parameters of nanostructures are estimated from the SEM images taken for different angles of 90$^o$ and 45$^o$ to the sample surface.

\textbf{Time-resolved spectroscopy.} Photoluminescence decay at room temperatures is investigated by a laser scanning confocal microscope MicroTime 100 (PicoQuant) equipped 
with an $\times$100 objective (NA=0.95) and 50-ps pulsed diode laser head ($\lambda$=405~nm), which implements the method of time-correlated single photon counting. 
All-measurement are made with laser pulses coming with the repetition rate 2500~kHz and fluence $F \approx 10 \mu J/cm^2$. 
Variations of the PL decay time from one sample to another are related to slight statistical deviations in conditions for each sample preparation and also related to inhomogeneity of each sample.

 \begin{figure*}
  \includegraphics[width=0.95\linewidth]{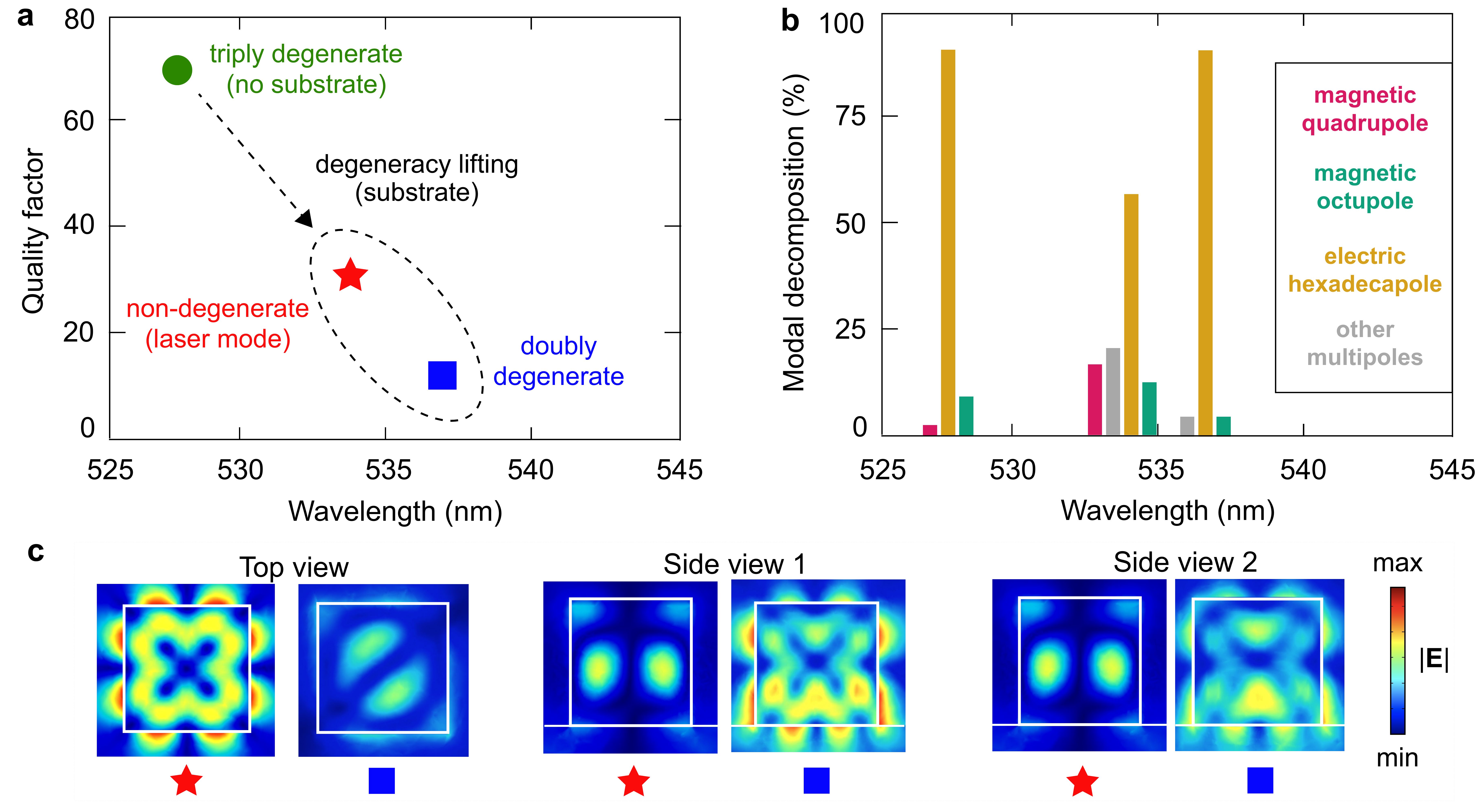}
  \caption{\textbf{Lasing modes of the smallest nanocube.} (a) Mode $Q$ factor vs. wavelength for the 420-nm CsPbBr$_3$ nanocube. A green circle marks a triply-degenerate mode of a nanocube. An arrow and a dashed oval show the effect of degeneracy lifting due to a substrate. Nanocube on a substrate supports a high-quality lasing mode (red star) and doubly-degenerate low-$Q$ mode (blue square). (b) Multipolar decomposition for the nanocube Mie eigenmodes. (c) Near-field profiles of the electric field for different modes.}
  \label{fig4}
\end{figure*}

\textbf{Dark-field spectroscopy.} Optical resonances of the perovskite nanocubes are studied by confocal dark-field optical spectroscopy. The nanocubes are excited at an oblique angle (65$^o$ with respect to the normal of the surface) by linearly polarized light from a halogen lamp (HL-2000-FHSA) through a weakly-focusing  objective (Mitutoyo M Plan Apo NIR, 10$\times$, NA = 0.28). Scattered light is collected from the top by an 50$\times$ objective (Mitutoyo M Plan APO NIR, NA = 0.42), sent to Horiba LabRam HR spectrometer and projected onto a thermoelectrically cooled charge-coupled device (CCD, Andor DU 420A-OE 325) with a 150-g/mm diffraction grating.

\textbf{Lasing measurements.}  To study lasing properties of individual nanocubes, we use frequency doubled Yb-doped femtosecond laser (TeMa, Avesta Project) yielding laser pulses at the wavelength of 524~nm, repetition rate 100~kHz after a Pockels cell, and pulse duration around 150~fs. The laser beam is focused onto the sample surface through the substrate at the normal incidence by a 10$\times$ objective (Mitutoyo M Plan APO NIR, NA = 0.28). Emission light is collected from the top by an 50$\times$ objective (Mitutoyo M Plan APO NIR, NA = 0.42), sent into Horiba LabRam HR spectrometer and projected onto a thermoelectrically cooled charge-coupled device (CCD, Andor DU 420A-OE 325). Lasing emission from individual nanocubes is studied by using a 1800-g/mm grating. All measurements are done in air at room temperature.

\textbf{FDTD simulations of the lasing threshold.}  Numerical calculations of the lasing threshold in perovskite nanoparticles are performed by using the finite-difference time-domain method with the use of a commercial software (Lumerical). The permittivity required for the estimation of the lasing threshold is introduced as a single Lorentzian with a negative oscillator strength $f_0$: $\varepsilon (\omega ) = {\varepsilon _0} + f_0 \omega _0^2(\omega _0^2 - {\omega ^2} - i\gamma \omega )^{-1}$, where $\varepsilon_0=6.25$ is the background permittivity of perovskite due to higher energy transitions, $\omega_0=2.31$ eV is the emission energy, and $\gamma=50$ meV is the emission linewidth. The system is excited with a set of randomly oriented electric dipole sources (to ensure coupling to all eigenmodes) with the perfect matched layer (PML) boundary conditions. The lasing threshold is then estimated as the minimum negative oscillator strength of the gain medium $f_{\rm th}$ (and the corresponding gain coefficient $-\text{Im }k$), at which the scattering spectrum diverges at a real-valued frequency.

\textbf{Eigenmode and scattering simulations.}  For numerical simulations of the eigenmode spectra and scattering spectra of the resonator, we use the finite-element method eigenmode solver and frequency domain solver in COMSOL Multiphysics, respectively. All calculations are realized for a single nanoresonator of a specific size in a homogeneous environment or on a semi-infinite substrate surrounded by PML mimicking an infinite region. The values for the refractive index for CsPbBr$_3$ are taken from experimentally measured data for perovskite thin films~\cite{tiguntseva2018tunable} and increased by 0.25 to consider the high quality material structure~\cite{leguy2016experimental} of fabricated monocrystalline cubes. The extinction coefficient is taken equal to zero for wavelengths longer than 530~nm, which allows to analyze the mode contribution in the lasing regime. For the scattering simulations, the excitation is a plane wave with polarization and incident angle as in the dark field spectroscopy experiment. The aperture of a collecting objective is NA=0.42, which reproduces the dark field spectroscopy experiment measurements.

\textbf{Multipolar decomposition.}
For characterization of the eigenmode spectrum, we apply the mode decomposition method over  irreducible  spherical multipoles~\cite{grahn2012electromagnetic}, characterized by both azimuthal ($m$) and orbital ($l$) indices. The decomposition is realized as a custom built-in routine for the eigenmode solver in COMSOL Multiphysics. For multipole classification summarized in Fig.~\ref{fig4}(b), we use the following notations: magnetic quadrupole ($l=2$), magnetic octupole ($l=3$), and electric hexadecapole ($l=4$).

{\bf Acknowledgements.} This work was supported by the Ministry of Education and Science of the Russian Federation (project 14.Y26.31.0010), for simulations and optical measurements, the Russian Science Foundation (project 19-73-30023), for synthesis and characterization, and the Strategic Fund of the Australian National University. The authors also acknowledge the Interdisciplinary Resource Center for Nanotechnology of Research Park of Saint Petersburg State University.

\bibliography{references}

\begin{thebibliography}{33}%
\makeatletter
\providecommand \@ifxundefined [1]{%
 \@ifx{#1\undefined}
}%
\providecommand \@ifnum [1]{%
 \ifnum #1\expandafter \@firstoftwo
 \else \expandafter \@secondoftwo
 \fi
}%
\providecommand \@ifx [1]{%
 \ifx #1\expandafter \@firstoftwo
 \else \expandafter \@secondoftwo
 \fi
}%
\providecommand \natexlab [1]{#1}%
\providecommand \enquote  [1]{``#1''}%
\providecommand \bibnamefont  [1]{#1}%
\providecommand \bibfnamefont [1]{#1}%
\providecommand \citenamefont [1]{#1}%
\providecommand \href@noop [0]{\@secondoftwo}%
\providecommand \href [0]{\begingroup \@sanitize@url \@href}%
\providecommand \@href[1]{\@@startlink{#1}\@@href}%
\providecommand \@@href[1]{\endgroup#1\@@endlink}%
\providecommand \@sanitize@url [0]{\catcode `\\12\catcode `\$12\catcode
  `\&12\catcode `\#12\catcode `\^12\catcode `\_12\catcode `\%12\relax}%
\providecommand \@@startlink[1]{}%
\providecommand \@@endlink[0]{}%
\providecommand \url  [0]{\begingroup\@sanitize@url \@url }%
\providecommand \@url [1]{\endgroup\@href {#1}{\urlprefix }}%
\providecommand \urlprefix  [0]{URL }%
\providecommand \Eprint [0]{\href }%
\providecommand \doibase [0]{http://dx.doi.org/}%
\providecommand \selectlanguage [0]{\@gobble}%
\providecommand \bibinfo  [0]{\@secondoftwo}%
\providecommand \bibfield  [0]{\@secondoftwo}%
\providecommand \translation [1]{[#1]}%
\providecommand \BibitemOpen [0]{}%
\providecommand \bibitemStop [0]{}%
\providecommand \bibitemNoStop [0]{.\EOS\space}%
\providecommand \EOS [0]{\spacefactor3000\relax}%
\providecommand \BibitemShut  [1]{\csname bibitem#1\endcsname}%
\let\auto@bib@innerbib\@empty
%</preamble>
\bibitem [{\citenamefont {Kuznetsov}\ \emph {et~al.}(2016)\citenamefont
  {Kuznetsov}, \citenamefont {Miroshnichenko}, \citenamefont {Brongersma},
  \citenamefont {Kivshar},\ and\ \citenamefont {Lukyanchuk}}]{kuznetsovreview}%
  \BibitemOpen
  \bibfield  {author} {\bibinfo {author} {\bibfnamefont {A.~I.}\ \bibnamefont
  {Kuznetsov}}, \bibinfo {author} {\bibfnamefont {A.~E.}\ \bibnamefont
  {Miroshnichenko}}, \bibinfo {author} {\bibfnamefont {M.~L.}\ \bibnamefont
  {Brongersma}}, \bibinfo {author} {\bibfnamefont {Y.~S.}\ \bibnamefont
  {Kivshar}}, \ and\ \bibinfo {author} {\bibfnamefont {B.}~\bibnamefont
  {Lukyanchuk}},\ }\href@noop {} {\bibfield  {journal} {\bibinfo  {journal}
  {Science}\ }\textbf {\bibinfo {volume} {354}},\ \bibinfo {pages} {aag2472}
  (\bibinfo {year} {2016})}\BibitemShut {NoStop}%
\bibitem [{\citenamefont {Rutckaia}\ \emph {et~al.}(2017)\citenamefont
  {Rutckaia}, \citenamefont {Heyroth}, \citenamefont {Novikov}, \citenamefont
  {Shaleev}, \citenamefont {Petrov},\ and\ \citenamefont
  {Schilling}}]{rutckaia2017quantum}%
  \BibitemOpen
  \bibfield  {author} {\bibinfo {author} {\bibfnamefont {V.}~\bibnamefont
  {Rutckaia}}, \bibinfo {author} {\bibfnamefont {F.}~\bibnamefont {Heyroth}},
  \bibinfo {author} {\bibfnamefont {A.}~\bibnamefont {Novikov}}, \bibinfo
  {author} {\bibfnamefont {M.}~\bibnamefont {Shaleev}}, \bibinfo {author}
  {\bibfnamefont {M.}~\bibnamefont {Petrov}}, \ and\ \bibinfo {author}
  {\bibfnamefont {J.}~\bibnamefont {Schilling}},\ }\href@noop {} {\bibfield
  {journal} {\bibinfo  {journal} {Nano Letters}\ }\textbf {\bibinfo {volume}
  {17}},\ \bibinfo {pages} {6886} (\bibinfo {year} {2017})}\BibitemShut
  {NoStop}%
\bibitem [{\citenamefont {Tiguntseva}\ \emph
  {et~al.}(2018{\natexlab{a}})\citenamefont {Tiguntseva}, \citenamefont
  {Zograf}, \citenamefont {Komissarenko}, \citenamefont {Zuev}, \citenamefont
  {Zakhidov}, \citenamefont {Makarov},\ and\ \citenamefont
  {Kivshar}}]{tiguntseva2018light}%
  \BibitemOpen
  \bibfield  {author} {\bibinfo {author} {\bibfnamefont {E.}~\bibnamefont
  {Tiguntseva}}, \bibinfo {author} {\bibfnamefont {G.~P.}\ \bibnamefont
  {Zograf}}, \bibinfo {author} {\bibfnamefont {F.~E.}\ \bibnamefont
  {Komissarenko}}, \bibinfo {author} {\bibfnamefont {D.~A.}\ \bibnamefont
  {Zuev}}, \bibinfo {author} {\bibfnamefont {A.~A.}\ \bibnamefont {Zakhidov}},
  \bibinfo {author} {\bibfnamefont {S.~V.}\ \bibnamefont {Makarov}}, \ and\
  \bibinfo {author} {\bibfnamefont {Y.~S.}\ \bibnamefont {Kivshar}},\
  }\href@noop {} {\bibfield  {journal} {\bibinfo  {journal} {Nano Letters}\
  }\textbf {\bibinfo {volume} {18}},\ \bibinfo {pages} {1185} (\bibinfo {year}
  {2018}{\natexlab{a}})}\BibitemShut {NoStop}%
\bibitem [{\citenamefont {Ha}\ \emph {et~al.}(2018)\citenamefont {Ha},
  \citenamefont {Fu}, \citenamefont {Emani}, \citenamefont {Pan}, \citenamefont
  {Bakker}, \citenamefont {Paniagua-Dom{\'\i}nguez},\ and\ \citenamefont
  {Kuznetsov}}]{ha2018directional}%
  \BibitemOpen
  \bibfield  {author} {\bibinfo {author} {\bibfnamefont {S.~T.}\ \bibnamefont
  {Ha}}, \bibinfo {author} {\bibfnamefont {Y.~H.}\ \bibnamefont {Fu}}, \bibinfo
  {author} {\bibfnamefont {N.~K.}\ \bibnamefont {Emani}}, \bibinfo {author}
  {\bibfnamefont {Z.}~\bibnamefont {Pan}}, \bibinfo {author} {\bibfnamefont
  {R.~M.}\ \bibnamefont {Bakker}}, \bibinfo {author} {\bibfnamefont
  {R.}~\bibnamefont {Paniagua-Dom{\'\i}nguez}}, \ and\ \bibinfo {author}
  {\bibfnamefont {A.~I.}\ \bibnamefont {Kuznetsov}},\ }\href@noop {} {\bibfield
   {journal} {\bibinfo  {journal} {Nature Nanotechnology}\ }\textbf {\bibinfo
  {volume} {13}},\ \bibinfo {pages} {1042} (\bibinfo {year}
  {2018})}\BibitemShut {NoStop}%
\bibitem [{\citenamefont {Kelso}\ \emph {et~al.}(2019)\citenamefont {Kelso},
  \citenamefont {Mahenderkar}, \citenamefont {Chen}, \citenamefont
  {Tubbesing},\ and\ \citenamefont {Switzer}}]{kelso2019spin}%
  \BibitemOpen
  \bibfield  {author} {\bibinfo {author} {\bibfnamefont {M.~V.}\ \bibnamefont
  {Kelso}}, \bibinfo {author} {\bibfnamefont {N.~K.}\ \bibnamefont
  {Mahenderkar}}, \bibinfo {author} {\bibfnamefont {Q.}~\bibnamefont {Chen}},
  \bibinfo {author} {\bibfnamefont {J.~Z.}\ \bibnamefont {Tubbesing}}, \ and\
  \bibinfo {author} {\bibfnamefont {J.~A.}\ \bibnamefont {Switzer}},\
  }\href@noop {} {\bibfield  {journal} {\bibinfo  {journal} {Science}\ }\textbf
  {\bibinfo {volume} {364}},\ \bibinfo {pages} {166} (\bibinfo {year}
  {2019})}\BibitemShut {NoStop}%
\bibitem [{\citenamefont {Shcherbakov}\ \emph {et~al.}(2014)\citenamefont
  {Shcherbakov}, \citenamefont {Neshev}, \citenamefont {Hopkins}, \citenamefont
  {Shorokhov}, \citenamefont {Staude}, \citenamefont {Melik-Gaykazyan},
  \citenamefont {Decker}, \citenamefont {Ezhov}, \citenamefont
  {Miroshnichenko}, \citenamefont {Brener}, \citenamefont {Fedyanin},\ and\
  \citenamefont {Kivshar}}]{shcherbakov2014enhanced}%
  \BibitemOpen
  \bibfield  {author} {\bibinfo {author} {\bibfnamefont {M.~R.}\ \bibnamefont
  {Shcherbakov}}, \bibinfo {author} {\bibfnamefont {D.~N.}\ \bibnamefont
  {Neshev}}, \bibinfo {author} {\bibfnamefont {B.}~\bibnamefont {Hopkins}},
  \bibinfo {author} {\bibfnamefont {A.~S.}\ \bibnamefont {Shorokhov}}, \bibinfo
  {author} {\bibfnamefont {I.}~\bibnamefont {Staude}}, \bibinfo {author}
  {\bibfnamefont {E.~V.}\ \bibnamefont {Melik-Gaykazyan}}, \bibinfo {author}
  {\bibfnamefont {M.}~\bibnamefont {Decker}}, \bibinfo {author} {\bibfnamefont
  {A.~A.}\ \bibnamefont {Ezhov}}, \bibinfo {author} {\bibfnamefont {A.~E.}\
  \bibnamefont {Miroshnichenko}}, \bibinfo {author} {\bibfnamefont
  {I.}~\bibnamefont {Brener}}, \bibinfo {author} {\bibfnamefont
  {A.}~\bibnamefont {Fedyanin}}, \ and\ \bibinfo {author} {\bibfnamefont
  {Y.}~\bibnamefont {Kivshar}},\ }\href@noop {} {\bibfield  {journal} {\bibinfo
   {journal} {Nano Letters}\ }\textbf {\bibinfo {volume} {14}},\ \bibinfo
  {pages} {6488} (\bibinfo {year} {2014})}\BibitemShut {NoStop}%
\bibitem [{\citenamefont {Kruk}\ \emph {et~al.}(2019)\citenamefont {Kruk},
  \citenamefont {Poddubny}, \citenamefont {Smirnova}, \citenamefont {Wang},
  \citenamefont {Slobozhanyuk}, \citenamefont {Shorokhov}, \citenamefont
  {Kravchenko}, \citenamefont {Luther-Davies},\ and\ \citenamefont
  {Kivshar}}]{kruk2019nonlinear}%
  \BibitemOpen
  \bibfield  {author} {\bibinfo {author} {\bibfnamefont {S.}~\bibnamefont
  {Kruk}}, \bibinfo {author} {\bibfnamefont {A.}~\bibnamefont {Poddubny}},
  \bibinfo {author} {\bibfnamefont {D.}~\bibnamefont {Smirnova}}, \bibinfo
  {author} {\bibfnamefont {L.}~\bibnamefont {Wang}}, \bibinfo {author}
  {\bibfnamefont {A.}~\bibnamefont {Slobozhanyuk}}, \bibinfo {author}
  {\bibfnamefont {A.}~\bibnamefont {Shorokhov}}, \bibinfo {author}
  {\bibfnamefont {I.}~\bibnamefont {Kravchenko}}, \bibinfo {author}
  {\bibfnamefont {B.}~\bibnamefont {Luther-Davies}}, \ and\ \bibinfo {author}
  {\bibfnamefont {Y.}~\bibnamefont {Kivshar}},\ }\href@noop {} {\bibfield
  {journal} {\bibinfo  {journal} {Nature Nanotechnology}\ }\textbf {\bibinfo
  {volume} {14}},\ \bibinfo {pages} {126} (\bibinfo {year} {2019})}\BibitemShut
  {NoStop}%
\bibitem [{\citenamefont {Dmitriev}\ \emph {et~al.}(2016)\citenamefont
  {Dmitriev}, \citenamefont {Baranov}, \citenamefont {Milichko}, \citenamefont
  {Makarov}, \citenamefont {Mukhin}, \citenamefont {Samusev}, \citenamefont
  {Krasnok}, \citenamefont {Belov},\ and\ \citenamefont
  {Kivshar}}]{dmitriev2016resonant}%
  \BibitemOpen
  \bibfield  {author} {\bibinfo {author} {\bibfnamefont {P.~A.}\ \bibnamefont
  {Dmitriev}}, \bibinfo {author} {\bibfnamefont {D.~G.}\ \bibnamefont
  {Baranov}}, \bibinfo {author} {\bibfnamefont {V.~A.}\ \bibnamefont
  {Milichko}}, \bibinfo {author} {\bibfnamefont {S.~V.}\ \bibnamefont
  {Makarov}}, \bibinfo {author} {\bibfnamefont {I.~S.}\ \bibnamefont {Mukhin}},
  \bibinfo {author} {\bibfnamefont {A.~K.}\ \bibnamefont {Samusev}}, \bibinfo
  {author} {\bibfnamefont {A.~E.}\ \bibnamefont {Krasnok}}, \bibinfo {author}
  {\bibfnamefont {P.~A.}\ \bibnamefont {Belov}}, \ and\ \bibinfo {author}
  {\bibfnamefont {Y.~S.}\ \bibnamefont {Kivshar}},\ }\href@noop {} {\bibfield
  {journal} {\bibinfo  {journal} {Nanoscale}\ }\textbf {\bibinfo {volume}
  {8}},\ \bibinfo {pages} {9721} (\bibinfo {year} {2016})}\BibitemShut
  {NoStop}%
\bibitem [{\citenamefont {Fratalocchi}\ \emph {et~al.}(2008)\citenamefont
  {Fratalocchi}, \citenamefont {Conti},\ and\ \citenamefont
  {Ruocco}}]{fratalocchi2008three}%
  \BibitemOpen
  \bibfield  {author} {\bibinfo {author} {\bibfnamefont {A.}~\bibnamefont
  {Fratalocchi}}, \bibinfo {author} {\bibfnamefont {C.}~\bibnamefont {Conti}},
  \ and\ \bibinfo {author} {\bibfnamefont {G.}~\bibnamefont {Ruocco}},\
  }\href@noop {} {\bibfield  {journal} {\bibinfo  {journal} {Physical Review
  A}\ }\textbf {\bibinfo {volume} {78}},\ \bibinfo {pages} {013806} (\bibinfo
  {year} {2008})}\BibitemShut {NoStop}%
\bibitem [{\citenamefont {Gongora}\ \emph {et~al.}(2017)\citenamefont
  {Gongora}, \citenamefont {Miroshnichenko}, \citenamefont {Kivshar},\ and\
  \citenamefont {Fratalocchi}}]{gongora2017anapole}%
  \BibitemOpen
  \bibfield  {author} {\bibinfo {author} {\bibfnamefont {J.~S.~T.}\
  \bibnamefont {Gongora}}, \bibinfo {author} {\bibfnamefont {A.~E.}\
  \bibnamefont {Miroshnichenko}}, \bibinfo {author} {\bibfnamefont {Y.~S.}\
  \bibnamefont {Kivshar}}, \ and\ \bibinfo {author} {\bibfnamefont
  {A.}~\bibnamefont {Fratalocchi}},\ }\href@noop {} {\bibfield  {journal}
  {\bibinfo  {journal} {Nature Communications}\ }\textbf {\bibinfo {volume}
  {8}},\ \bibinfo {pages} {15535} (\bibinfo {year} {2017})}\BibitemShut
  {NoStop}%
\bibitem [{\citenamefont {Sutherland}\ and\ \citenamefont
  {Sargent}(2016)}]{sutherland2016perovskite}%
  \BibitemOpen
  \bibfield  {author} {\bibinfo {author} {\bibfnamefont {B.~R.}\ \bibnamefont
  {Sutherland}}\ and\ \bibinfo {author} {\bibfnamefont {E.~H.}\ \bibnamefont
  {Sargent}},\ }\href@noop {} {\bibfield  {journal} {\bibinfo  {journal}
  {Nature Photonics}\ }\textbf {\bibinfo {volume} {10}},\ \bibinfo {pages}
  {295} (\bibinfo {year} {2016})}\BibitemShut {NoStop}%
\bibitem [{\citenamefont {Makarov}\ \emph {et~al.}(2019)\citenamefont
  {Makarov}, \citenamefont {Furasova}, \citenamefont {Tiguntseva},
  \citenamefont {Hemmetter}, \citenamefont {Berestennikov}, \citenamefont
  {Pushkarev}, \citenamefont {Zakhidov},\ and\ \citenamefont
  {Kivshar}}]{makarov2019halide}%
  \BibitemOpen
  \bibfield  {author} {\bibinfo {author} {\bibfnamefont {S.}~\bibnamefont
  {Makarov}}, \bibinfo {author} {\bibfnamefont {A.}~\bibnamefont {Furasova}},
  \bibinfo {author} {\bibfnamefont {E.}~\bibnamefont {Tiguntseva}}, \bibinfo
  {author} {\bibfnamefont {A.}~\bibnamefont {Hemmetter}}, \bibinfo {author}
  {\bibfnamefont {A.}~\bibnamefont {Berestennikov}}, \bibinfo {author}
  {\bibfnamefont {A.}~\bibnamefont {Pushkarev}}, \bibinfo {author}
  {\bibfnamefont {A.}~\bibnamefont {Zakhidov}}, \ and\ \bibinfo {author}
  {\bibfnamefont {Y.}~\bibnamefont {Kivshar}},\ }\href@noop {} {\bibfield
  {journal} {\bibinfo  {journal} {Advanced Optical Materials}\ }\textbf
  {\bibinfo {volume} {7}},\ \bibinfo {pages} {1800784} (\bibinfo {year}
  {2019})}\BibitemShut {NoStop}%
\bibitem [{\citenamefont {Gao}\ \emph {et~al.}(2018)\citenamefont {Gao},
  \citenamefont {Huang}, \citenamefont {Hao}, \citenamefont {Sun},
  \citenamefont {Zhang}, \citenamefont {Zhang}, \citenamefont {Duan},
  \citenamefont {Wang}, \citenamefont {Jin},\ and\ \citenamefont
  {Zhang}}]{gao2018lead}%
  \BibitemOpen
  \bibfield  {author} {\bibinfo {author} {\bibfnamefont {Y.}~\bibnamefont
  {Gao}}, \bibinfo {author} {\bibfnamefont {C.}~\bibnamefont {Huang}}, \bibinfo
  {author} {\bibfnamefont {C.}~\bibnamefont {Hao}}, \bibinfo {author}
  {\bibfnamefont {S.}~\bibnamefont {Sun}}, \bibinfo {author} {\bibfnamefont
  {L.}~\bibnamefont {Zhang}}, \bibinfo {author} {\bibfnamefont
  {C.}~\bibnamefont {Zhang}}, \bibinfo {author} {\bibfnamefont
  {Z.}~\bibnamefont {Duan}}, \bibinfo {author} {\bibfnamefont {K.}~\bibnamefont
  {Wang}}, \bibinfo {author} {\bibfnamefont {Z.}~\bibnamefont {Jin}}, \ and\
  \bibinfo {author} {\bibfnamefont {N.}~\bibnamefont {Zhang}},\ }\href@noop {}
  {\bibfield  {journal} {\bibinfo  {journal} {ACS Nano}\ }\textbf {\bibinfo
  {volume} {12}},\ \bibinfo {pages} {8847} (\bibinfo {year}
  {2018})}\BibitemShut {NoStop}%
\bibitem [{\citenamefont {Su}\ \emph {et~al.}(2018)\citenamefont {Su},
  \citenamefont {Wang}, \citenamefont {Zhao}, \citenamefont {Xing},
  \citenamefont {Zhao}, \citenamefont {Diederichs}, \citenamefont {Liew},\ and\
  \citenamefont {Xiong}}]{su2018room}%
  \BibitemOpen
  \bibfield  {author} {\bibinfo {author} {\bibfnamefont {R.}~\bibnamefont
  {Su}}, \bibinfo {author} {\bibfnamefont {J.}~\bibnamefont {Wang}}, \bibinfo
  {author} {\bibfnamefont {J.}~\bibnamefont {Zhao}}, \bibinfo {author}
  {\bibfnamefont {J.}~\bibnamefont {Xing}}, \bibinfo {author} {\bibfnamefont
  {W.}~\bibnamefont {Zhao}}, \bibinfo {author} {\bibfnamefont {C.}~\bibnamefont
  {Diederichs}}, \bibinfo {author} {\bibfnamefont {T.~C.}\ \bibnamefont
  {Liew}}, \ and\ \bibinfo {author} {\bibfnamefont {Q.}~\bibnamefont {Xiong}},\
  }\href@noop {} {\bibfield  {journal} {\bibinfo  {journal} {Science Advances}\
  }\textbf {\bibinfo {volume} {4}},\ \bibinfo {pages} {eaau0244} (\bibinfo
  {year} {2018})}\BibitemShut {NoStop}%
\bibitem [{\citenamefont {Zhu}\ \emph {et~al.}(2015)\citenamefont {Zhu},
  \citenamefont {Fu}, \citenamefont {Meng}, \citenamefont {Wu}, \citenamefont
  {Gong}, \citenamefont {Ding}, \citenamefont {Gustafsson}, \citenamefont
  {Trinh}, \citenamefont {Jin},\ and\ \citenamefont {Zhu}}]{zhu2015lead}%
  \BibitemOpen
  \bibfield  {author} {\bibinfo {author} {\bibfnamefont {H.}~\bibnamefont
  {Zhu}}, \bibinfo {author} {\bibfnamefont {Y.}~\bibnamefont {Fu}}, \bibinfo
  {author} {\bibfnamefont {F.}~\bibnamefont {Meng}}, \bibinfo {author}
  {\bibfnamefont {X.}~\bibnamefont {Wu}}, \bibinfo {author} {\bibfnamefont
  {Z.}~\bibnamefont {Gong}}, \bibinfo {author} {\bibfnamefont {Q.}~\bibnamefont
  {Ding}}, \bibinfo {author} {\bibfnamefont {M.~V.}\ \bibnamefont
  {Gustafsson}}, \bibinfo {author} {\bibfnamefont {M.~T.}\ \bibnamefont
  {Trinh}}, \bibinfo {author} {\bibfnamefont {S.}~\bibnamefont {Jin}}, \ and\
  \bibinfo {author} {\bibfnamefont {X.}~\bibnamefont {Zhu}},\ }\href@noop {}
  {\bibfield  {journal} {\bibinfo  {journal} {Nature Materials}\ }\textbf
  {\bibinfo {volume} {14}},\ \bibinfo {pages} {636} (\bibinfo {year}
  {2015})}\BibitemShut {NoStop}%
\bibitem [{\citenamefont {Xing}\ \emph {et~al.}(2015)\citenamefont {Xing},
  \citenamefont {Liu}, \citenamefont {Zhang}, \citenamefont {Ha}, \citenamefont
  {Yuan}, \citenamefont {Shen}, \citenamefont {Sum},\ and\ \citenamefont
  {Xiong}}]{xing2015vapor}%
  \BibitemOpen
  \bibfield  {author} {\bibinfo {author} {\bibfnamefont {J.}~\bibnamefont
  {Xing}}, \bibinfo {author} {\bibfnamefont {X.~F.}\ \bibnamefont {Liu}},
  \bibinfo {author} {\bibfnamefont {Q.}~\bibnamefont {Zhang}}, \bibinfo
  {author} {\bibfnamefont {S.~T.}\ \bibnamefont {Ha}}, \bibinfo {author}
  {\bibfnamefont {Y.~W.}\ \bibnamefont {Yuan}}, \bibinfo {author}
  {\bibfnamefont {C.}~\bibnamefont {Shen}}, \bibinfo {author} {\bibfnamefont
  {T.~C.}\ \bibnamefont {Sum}}, \ and\ \bibinfo {author} {\bibfnamefont
  {Q.}~\bibnamefont {Xiong}},\ }\href@noop {} {\bibfield  {journal} {\bibinfo
  {journal} {Nano Letters}\ }\textbf {\bibinfo {volume} {15}},\ \bibinfo
  {pages} {4571} (\bibinfo {year} {2015})}\BibitemShut {NoStop}%
\bibitem [{\citenamefont {Shang}\ \emph {et~al.}(2018)\citenamefont {Shang},
  \citenamefont {Zhang}, \citenamefont {Liu}, \citenamefont {Chen},
  \citenamefont {Yang}, \citenamefont {Li}, \citenamefont {Li}, \citenamefont
  {Zhang}, \citenamefont {Xiong},\ and\ \citenamefont
  {Liu}}]{shang2018surface}%
  \BibitemOpen
  \bibfield  {author} {\bibinfo {author} {\bibfnamefont {Q.}~\bibnamefont
  {Shang}}, \bibinfo {author} {\bibfnamefont {S.}~\bibnamefont {Zhang}},
  \bibinfo {author} {\bibfnamefont {Z.}~\bibnamefont {Liu}}, \bibinfo {author}
  {\bibfnamefont {J.}~\bibnamefont {Chen}}, \bibinfo {author} {\bibfnamefont
  {P.}~\bibnamefont {Yang}}, \bibinfo {author} {\bibfnamefont {C.}~\bibnamefont
  {Li}}, \bibinfo {author} {\bibfnamefont {W.}~\bibnamefont {Li}}, \bibinfo
  {author} {\bibfnamefont {Y.}~\bibnamefont {Zhang}}, \bibinfo {author}
  {\bibfnamefont {Q.}~\bibnamefont {Xiong}}, \ and\ \bibinfo {author}
  {\bibfnamefont {X.}~\bibnamefont {Liu}},\ }\href@noop {} {\bibfield
  {journal} {\bibinfo  {journal} {Nano Letters}\ }\textbf {\bibinfo {volume}
  {18}},\ \bibinfo {pages} {3335} (\bibinfo {year} {2018})}\BibitemShut
  {NoStop}%
\bibitem [{\citenamefont {Zhang}\ \emph {et~al.}(2014)\citenamefont {Zhang},
  \citenamefont {Ha}, \citenamefont {Liu}, \citenamefont {Sum},\ and\
  \citenamefont {Xiong}}]{zhang2014room}%
  \BibitemOpen
  \bibfield  {author} {\bibinfo {author} {\bibfnamefont {Q.}~\bibnamefont
  {Zhang}}, \bibinfo {author} {\bibfnamefont {S.~T.}\ \bibnamefont {Ha}},
  \bibinfo {author} {\bibfnamefont {X.}~\bibnamefont {Liu}}, \bibinfo {author}
  {\bibfnamefont {T.~C.}\ \bibnamefont {Sum}}, \ and\ \bibinfo {author}
  {\bibfnamefont {Q.}~\bibnamefont {Xiong}},\ }\href@noop {} {\bibfield
  {journal} {\bibinfo  {journal} {Nano Letters}\ }\textbf {\bibinfo {volume}
  {14}},\ \bibinfo {pages} {5995} (\bibinfo {year} {2014})}\BibitemShut
  {NoStop}%
\bibitem [{\citenamefont {Zhang}\ \emph {et~al.}(2016)\citenamefont {Zhang},
  \citenamefont {Su}, \citenamefont {Liu}, \citenamefont {Xing}, \citenamefont
  {Sum},\ and\ \citenamefont {Xiong}}]{zhang2016high}%
  \BibitemOpen
  \bibfield  {author} {\bibinfo {author} {\bibfnamefont {Q.}~\bibnamefont
  {Zhang}}, \bibinfo {author} {\bibfnamefont {R.}~\bibnamefont {Su}}, \bibinfo
  {author} {\bibfnamefont {X.}~\bibnamefont {Liu}}, \bibinfo {author}
  {\bibfnamefont {J.}~\bibnamefont {Xing}}, \bibinfo {author} {\bibfnamefont
  {T.~C.}\ \bibnamefont {Sum}}, \ and\ \bibinfo {author} {\bibfnamefont
  {Q.}~\bibnamefont {Xiong}},\ }\href@noop {} {\bibfield  {journal} {\bibinfo
  {journal} {Advanced Functional Materials}\ }\textbf {\bibinfo {volume}
  {26}},\ \bibinfo {pages} {6238} (\bibinfo {year} {2016})}\BibitemShut
  {NoStop}%
\bibitem [{\citenamefont {Su}\ \emph {et~al.}(2017)\citenamefont {Su},
  \citenamefont {Diederichs}, \citenamefont {Wang}, \citenamefont {Liew},
  \citenamefont {Zhao}, \citenamefont {Liu}, \citenamefont {Xu}, \citenamefont
  {Chen},\ and\ \citenamefont {Xiong}}]{su2017room}%
  \BibitemOpen
  \bibfield  {author} {\bibinfo {author} {\bibfnamefont {R.}~\bibnamefont
  {Su}}, \bibinfo {author} {\bibfnamefont {C.}~\bibnamefont {Diederichs}},
  \bibinfo {author} {\bibfnamefont {J.}~\bibnamefont {Wang}}, \bibinfo {author}
  {\bibfnamefont {T.~C.}\ \bibnamefont {Liew}}, \bibinfo {author}
  {\bibfnamefont {J.}~\bibnamefont {Zhao}}, \bibinfo {author} {\bibfnamefont
  {S.}~\bibnamefont {Liu}}, \bibinfo {author} {\bibfnamefont {W.}~\bibnamefont
  {Xu}}, \bibinfo {author} {\bibfnamefont {Z.}~\bibnamefont {Chen}}, \ and\
  \bibinfo {author} {\bibfnamefont {Q.}~\bibnamefont {Xiong}},\ }\href@noop {}
  {\bibfield  {journal} {\bibinfo  {journal} {Nano Letters}\ }\textbf {\bibinfo
  {volume} {17}},\ \bibinfo {pages} {3982} (\bibinfo {year}
  {2017})}\BibitemShut {NoStop}%
\bibitem [{\citenamefont {Tang}\ \emph {et~al.}(2017)\citenamefont {Tang},
  \citenamefont {Dong}, \citenamefont {Sun}, \citenamefont {Zheng},
  \citenamefont {Wang}, \citenamefont {Sun}, \citenamefont {Jiang},
  \citenamefont {Pan},\ and\ \citenamefont {Zhang}}]{tang2017single}%
  \BibitemOpen
  \bibfield  {author} {\bibinfo {author} {\bibfnamefont {B.}~\bibnamefont
  {Tang}}, \bibinfo {author} {\bibfnamefont {H.}~\bibnamefont {Dong}}, \bibinfo
  {author} {\bibfnamefont {L.}~\bibnamefont {Sun}}, \bibinfo {author}
  {\bibfnamefont {W.}~\bibnamefont {Zheng}}, \bibinfo {author} {\bibfnamefont
  {Q.}~\bibnamefont {Wang}}, \bibinfo {author} {\bibfnamefont {F.}~\bibnamefont
  {Sun}}, \bibinfo {author} {\bibfnamefont {X.}~\bibnamefont {Jiang}}, \bibinfo
  {author} {\bibfnamefont {A.}~\bibnamefont {Pan}}, \ and\ \bibinfo {author}
  {\bibfnamefont {L.}~\bibnamefont {Zhang}},\ }\href@noop {} {\bibfield
  {journal} {\bibinfo  {journal} {ACS Nano}\ }\textbf {\bibinfo {volume}
  {11}},\ \bibinfo {pages} {10681} (\bibinfo {year} {2017})}\BibitemShut
  {NoStop}%
\bibitem [{\citenamefont {Fu}\ \emph {et~al.}(2016)\citenamefont {Fu},
  \citenamefont {Zhu}, \citenamefont {Schrader}, \citenamefont {Liang},
  \citenamefont {Ding}, \citenamefont {Joshi}, \citenamefont {Hwang},
  \citenamefont {Zhu},\ and\ \citenamefont {Jin}}]{fu2016nanowire}%
  \BibitemOpen
  \bibfield  {author} {\bibinfo {author} {\bibfnamefont {Y.}~\bibnamefont
  {Fu}}, \bibinfo {author} {\bibfnamefont {H.}~\bibnamefont {Zhu}}, \bibinfo
  {author} {\bibfnamefont {A.~W.}\ \bibnamefont {Schrader}}, \bibinfo {author}
  {\bibfnamefont {D.}~\bibnamefont {Liang}}, \bibinfo {author} {\bibfnamefont
  {Q.}~\bibnamefont {Ding}}, \bibinfo {author} {\bibfnamefont {P.}~\bibnamefont
  {Joshi}}, \bibinfo {author} {\bibfnamefont {L.}~\bibnamefont {Hwang}},
  \bibinfo {author} {\bibfnamefont {X.}~\bibnamefont {Zhu}}, \ and\ \bibinfo
  {author} {\bibfnamefont {S.}~\bibnamefont {Jin}},\ }\href@noop {} {\bibfield
  {journal} {\bibinfo  {journal} {Nano Letters}\ }\textbf {\bibinfo {volume}
  {16}},\ \bibinfo {pages} {1000} (\bibinfo {year} {2016})}\BibitemShut
  {NoStop}%
\bibitem [{\citenamefont {Eaton}\ \emph {et~al.}(2016)\citenamefont {Eaton},
  \citenamefont {Lai}, \citenamefont {Gibson}, \citenamefont {Wong},
  \citenamefont {Dou}, \citenamefont {Ma}, \citenamefont {Wang}, \citenamefont
  {Leone},\ and\ \citenamefont {Yang}}]{eaton2016lasing}%
  \BibitemOpen
  \bibfield  {author} {\bibinfo {author} {\bibfnamefont {S.~W.}\ \bibnamefont
  {Eaton}}, \bibinfo {author} {\bibfnamefont {M.}~\bibnamefont {Lai}}, \bibinfo
  {author} {\bibfnamefont {N.~A.}\ \bibnamefont {Gibson}}, \bibinfo {author}
  {\bibfnamefont {A.~B.}\ \bibnamefont {Wong}}, \bibinfo {author}
  {\bibfnamefont {L.}~\bibnamefont {Dou}}, \bibinfo {author} {\bibfnamefont
  {J.}~\bibnamefont {Ma}}, \bibinfo {author} {\bibfnamefont {L.-W.}\
  \bibnamefont {Wang}}, \bibinfo {author} {\bibfnamefont {S.~R.}\ \bibnamefont
  {Leone}}, \ and\ \bibinfo {author} {\bibfnamefont {P.}~\bibnamefont {Yang}},\
  }\href@noop {} {\bibfield  {journal} {\bibinfo  {journal} {Proceedings of the
  National Academy of Sciences}\ }\textbf {\bibinfo {volume} {113}},\ \bibinfo
  {pages} {1993} (\bibinfo {year} {2016})}\BibitemShut {NoStop}%
\bibitem [{\citenamefont {Park}\ \emph {et~al.}(2016)\citenamefont {Park},
  \citenamefont {Lee}, \citenamefont {Kim}, \citenamefont {Han}, \citenamefont
  {Jang}, \citenamefont {Jeong}, \citenamefont {Park},\ and\ \citenamefont
  {Song}}]{park2016light}%
  \BibitemOpen
  \bibfield  {author} {\bibinfo {author} {\bibfnamefont {K.}~\bibnamefont
  {Park}}, \bibinfo {author} {\bibfnamefont {J.~W.}\ \bibnamefont {Lee}},
  \bibinfo {author} {\bibfnamefont {J.~D.}\ \bibnamefont {Kim}}, \bibinfo
  {author} {\bibfnamefont {N.~S.}\ \bibnamefont {Han}}, \bibinfo {author}
  {\bibfnamefont {D.~M.}\ \bibnamefont {Jang}}, \bibinfo {author}
  {\bibfnamefont {S.}~\bibnamefont {Jeong}}, \bibinfo {author} {\bibfnamefont
  {J.}~\bibnamefont {Park}}, \ and\ \bibinfo {author} {\bibfnamefont {J.~K.}\
  \bibnamefont {Song}},\ }\href@noop {} {\bibfield  {journal} {\bibinfo
  {journal} {The Journal of Physical Chemistry Letters}\ }\textbf {\bibinfo
  {volume} {7}},\ \bibinfo {pages} {3703} (\bibinfo {year} {2016})}\BibitemShut
  {NoStop}%
\bibitem [{\citenamefont {Zhou}\ \emph {et~al.}(2018)\citenamefont {Zhou},
  \citenamefont {Dong}, \citenamefont {Jiang}, \citenamefont {Zheng},
  \citenamefont {Sun}, \citenamefont {Zhao}, \citenamefont {Tang},
  \citenamefont {Pan},\ and\ \citenamefont {Zhang}}]{zhou2018single}%
  \BibitemOpen
  \bibfield  {author} {\bibinfo {author} {\bibfnamefont {B.}~\bibnamefont
  {Zhou}}, \bibinfo {author} {\bibfnamefont {H.}~\bibnamefont {Dong}}, \bibinfo
  {author} {\bibfnamefont {M.}~\bibnamefont {Jiang}}, \bibinfo {author}
  {\bibfnamefont {W.}~\bibnamefont {Zheng}}, \bibinfo {author} {\bibfnamefont
  {L.}~\bibnamefont {Sun}}, \bibinfo {author} {\bibfnamefont {B.}~\bibnamefont
  {Zhao}}, \bibinfo {author} {\bibfnamefont {B.}~\bibnamefont {Tang}}, \bibinfo
  {author} {\bibfnamefont {A.}~\bibnamefont {Pan}}, \ and\ \bibinfo {author}
  {\bibfnamefont {L.}~\bibnamefont {Zhang}},\ }\href@noop {} {\bibfield
  {journal} {\bibinfo  {journal} {Journal of Materials Chemistry C}\ }\textbf
  {\bibinfo {volume} {6}},\ \bibinfo {pages} {11740} (\bibinfo {year}
  {2018})}\BibitemShut {NoStop}%
\bibitem [{\citenamefont {Brandt}\ \emph {et~al.}(2017)\citenamefont {Brandt},
  \citenamefont {Poindexter}, \citenamefont {Gorai}, \citenamefont {Kurchin},
  \citenamefont {Hoye}, \citenamefont {Nienhaus}, \citenamefont {Wilson},
  \citenamefont {Polizzotti}, \citenamefont {Sereika},\ and\ \citenamefont
  {Zaltauskas}}]{brandt2017searching}%
  \BibitemOpen
  \bibfield  {author} {\bibinfo {author} {\bibfnamefont {R.~E.}\ \bibnamefont
  {Brandt}}, \bibinfo {author} {\bibfnamefont {J.~R.}\ \bibnamefont
  {Poindexter}}, \bibinfo {author} {\bibfnamefont {P.}~\bibnamefont {Gorai}},
  \bibinfo {author} {\bibfnamefont {R.~C.}\ \bibnamefont {Kurchin}}, \bibinfo
  {author} {\bibfnamefont {R.~L.}\ \bibnamefont {Hoye}}, \bibinfo {author}
  {\bibfnamefont {L.}~\bibnamefont {Nienhaus}}, \bibinfo {author}
  {\bibfnamefont {M.~W.}\ \bibnamefont {Wilson}}, \bibinfo {author}
  {\bibfnamefont {J.~A.}\ \bibnamefont {Polizzotti}}, \bibinfo {author}
  {\bibfnamefont {R.}~\bibnamefont {Sereika}}, \ and\ \bibinfo {author}
  {\bibfnamefont {R.}~\bibnamefont {Zaltauskas}},\ }\href@noop {} {\bibfield
  {journal} {\bibinfo  {journal} {Chemistry of Materials}\ }\textbf {\bibinfo
  {volume} {29}},\ \bibinfo {pages} {4667} (\bibinfo {year}
  {2017})}\BibitemShut {NoStop}%
\bibitem [{\citenamefont {Gu}\ \emph {et~al.}(2013)\citenamefont {Gu},
  \citenamefont {Slutsky}, \citenamefont {Vallini}, \citenamefont {Smalley},
  \citenamefont {Nezhad}, \citenamefont {Frateschi},\ and\ \citenamefont
  {Fainman}}]{gu2013purcell}%
  \BibitemOpen
  \bibfield  {author} {\bibinfo {author} {\bibfnamefont {Q.}~\bibnamefont
  {Gu}}, \bibinfo {author} {\bibfnamefont {B.}~\bibnamefont {Slutsky}},
  \bibinfo {author} {\bibfnamefont {F.}~\bibnamefont {Vallini}}, \bibinfo
  {author} {\bibfnamefont {J.~S.}\ \bibnamefont {Smalley}}, \bibinfo {author}
  {\bibfnamefont {M.~P.}\ \bibnamefont {Nezhad}}, \bibinfo {author}
  {\bibfnamefont {N.~C.}\ \bibnamefont {Frateschi}}, \ and\ \bibinfo {author}
  {\bibfnamefont {Y.}~\bibnamefont {Fainman}},\ }\href@noop {} {\bibfield
  {journal} {\bibinfo  {journal} {Optics Express}\ }\textbf {\bibinfo {volume}
  {21}},\ \bibinfo {pages} {15603} (\bibinfo {year} {2013})}\BibitemShut
  {NoStop}%
\bibitem [{\citenamefont {Ma}\ and\ \citenamefont
  {Oulton}(2019)}]{ma2019applications}%
  \BibitemOpen
  \bibfield  {author} {\bibinfo {author} {\bibfnamefont {R.-M.}\ \bibnamefont
  {Ma}}\ and\ \bibinfo {author} {\bibfnamefont {R.~F.}\ \bibnamefont
  {Oulton}},\ }\href@noop {} {\bibfield  {journal} {\bibinfo  {journal} {Nature
  Nanotechnology}\ }\textbf {\bibinfo {volume} {14}},\ \bibinfo {pages} {12}
  (\bibinfo {year} {2019})}\BibitemShut {NoStop}%
\bibitem [{\citenamefont {Ge}\ \emph {et~al.}(2010)\citenamefont {Ge},
  \citenamefont {Chong},\ and\ \citenamefont {Stone}}]{PhysRevA.82.063824}%
  \BibitemOpen
  \bibfield  {author} {\bibinfo {author} {\bibfnamefont {L.}~\bibnamefont
  {Ge}}, \bibinfo {author} {\bibfnamefont {Y.~D.}\ \bibnamefont {Chong}}, \
  and\ \bibinfo {author} {\bibfnamefont {A.~D.}\ \bibnamefont {Stone}},\
  }\href@noop {} {\bibfield  {journal} {\bibinfo  {journal} {Physical Review
  A}\ }\textbf {\bibinfo {volume} {82}},\ \bibinfo {pages} {063824} (\bibinfo
  {year} {2010})}\BibitemShut {NoStop}%
\bibitem [{\citenamefont {Liu}\ \emph {et~al.}(2018)\citenamefont {Liu},
  \citenamefont {Yang}, \citenamefont {Du}, \citenamefont {Hu}, \citenamefont
  {Shi}, \citenamefont {Zhang}, \citenamefont {Liu}, \citenamefont {Tang},
  \citenamefont {Leng},\ and\ \citenamefont {Li}}]{liu2018robust}%
  \BibitemOpen
  \bibfield  {author} {\bibinfo {author} {\bibfnamefont {Z.}~\bibnamefont
  {Liu}}, \bibinfo {author} {\bibfnamefont {J.}~\bibnamefont {Yang}}, \bibinfo
  {author} {\bibfnamefont {J.}~\bibnamefont {Du}}, \bibinfo {author}
  {\bibfnamefont {Z.}~\bibnamefont {Hu}}, \bibinfo {author} {\bibfnamefont
  {T.}~\bibnamefont {Shi}}, \bibinfo {author} {\bibfnamefont {Z.}~\bibnamefont
  {Zhang}}, \bibinfo {author} {\bibfnamefont {Y.}~\bibnamefont {Liu}}, \bibinfo
  {author} {\bibfnamefont {X.}~\bibnamefont {Tang}}, \bibinfo {author}
  {\bibfnamefont {Y.}~\bibnamefont {Leng}}, \ and\ \bibinfo {author}
  {\bibfnamefont {R.}~\bibnamefont {Li}},\ }\href@noop {} {\bibfield  {journal}
  {\bibinfo  {journal} {ACS Nano}\ }\textbf {\bibinfo {volume} {12}},\ \bibinfo
  {pages} {5923} (\bibinfo {year} {2018})}\BibitemShut {NoStop}%
\bibitem [{\citenamefont {Tiguntseva}\ \emph
  {et~al.}(2018{\natexlab{b}})\citenamefont {Tiguntseva}, \citenamefont
  {Baranov}, \citenamefont {Pushkarev}, \citenamefont {Munkhbat}, \citenamefont
  {Komissarenko}, \citenamefont {Franckevicius}, \citenamefont {Zakhidov},
  \citenamefont {Shegai}, \citenamefont {Kivshar},\ and\ \citenamefont
  {Makarov}}]{tiguntseva2018tunable}%
  \BibitemOpen
  \bibfield  {author} {\bibinfo {author} {\bibfnamefont {E.}~\bibnamefont
  {Tiguntseva}}, \bibinfo {author} {\bibfnamefont {D.~G.}\ \bibnamefont
  {Baranov}}, \bibinfo {author} {\bibfnamefont {A.}~\bibnamefont {Pushkarev}},
  \bibinfo {author} {\bibfnamefont {B.}~\bibnamefont {Munkhbat}}, \bibinfo
  {author} {\bibfnamefont {F.~E.}\ \bibnamefont {Komissarenko}}, \bibinfo
  {author} {\bibfnamefont {M.}~\bibnamefont {Franckevicius}}, \bibinfo {author}
  {\bibfnamefont {A.~A.}\ \bibnamefont {Zakhidov}}, \bibinfo {author}
  {\bibfnamefont {T.}~\bibnamefont {Shegai}}, \bibinfo {author} {\bibfnamefont
  {Y.~S.}\ \bibnamefont {Kivshar}}, \ and\ \bibinfo {author} {\bibfnamefont
  {S.~V.}\ \bibnamefont {Makarov}},\ }\href@noop {} {\bibfield  {journal}
  {\bibinfo  {journal} {Nano Letters}\ }\textbf {\bibinfo {volume} {18}},\
  \bibinfo {pages} {5522} (\bibinfo {year} {2018}{\natexlab{b}})}\BibitemShut
  {NoStop}%
\bibitem [{\citenamefont {Leguy}\ \emph {et~al.}(2016)\citenamefont {Leguy},
  \citenamefont {Azarhoosh}, \citenamefont {Alonso}, \citenamefont
  {Campoy-Quiles}, \citenamefont {Weber}, \citenamefont {Yao}, \citenamefont
  {Bryant}, \citenamefont {Weller}, \citenamefont {Nelson},\ and\ \citenamefont
  {Walsh}}]{leguy2016experimental}%
  \BibitemOpen
  \bibfield  {author} {\bibinfo {author} {\bibfnamefont {A.~M.}\ \bibnamefont
  {Leguy}}, \bibinfo {author} {\bibfnamefont {P.}~\bibnamefont {Azarhoosh}},
  \bibinfo {author} {\bibfnamefont {M.~I.}\ \bibnamefont {Alonso}}, \bibinfo
  {author} {\bibfnamefont {M.}~\bibnamefont {Campoy-Quiles}}, \bibinfo {author}
  {\bibfnamefont {O.~J.}\ \bibnamefont {Weber}}, \bibinfo {author}
  {\bibfnamefont {J.}~\bibnamefont {Yao}}, \bibinfo {author} {\bibfnamefont
  {D.}~\bibnamefont {Bryant}}, \bibinfo {author} {\bibfnamefont {M.~T.}\
  \bibnamefont {Weller}}, \bibinfo {author} {\bibfnamefont {J.}~\bibnamefont
  {Nelson}}, \ and\ \bibinfo {author} {\bibfnamefont {A.}~\bibnamefont
  {Walsh}},\ }\href@noop {} {\bibfield  {journal} {\bibinfo  {journal}
  {Nanoscale}\ }\textbf {\bibinfo {volume} {8}},\ \bibinfo {pages} {6317}
  (\bibinfo {year} {2016})}\BibitemShut {NoStop}%
\bibitem [{\citenamefont {Grahn}\ \emph {et~al.}(2012)\citenamefont {Grahn},
  \citenamefont {Shevchenko},\ and\ \citenamefont
  {Kaivola}}]{grahn2012electromagnetic}%
  \BibitemOpen
  \bibfield  {author} {\bibinfo {author} {\bibfnamefont {P.}~\bibnamefont
  {Grahn}}, \bibinfo {author} {\bibfnamefont {A.}~\bibnamefont {Shevchenko}}, \
  and\ \bibinfo {author} {\bibfnamefont {M.}~\bibnamefont {Kaivola}},\
  }\href@noop {} {\bibfield  {journal} {\bibinfo  {journal} {New Journal of
  Physics}\ }\textbf {\bibinfo {volume} {14}},\ \bibinfo {pages} {093033}
  (\bibinfo {year} {2012})}\BibitemShut {NoStop}%
\end{thebibliography}%

\end{document}